\documentclass[sigplan,screen]{acmart}

\copyrightyear{2025}
\acmYear{2025}
\setcopyright{cc}
\setcctype{CC-BY}
\acmConference[ASPLOS '25]{Proceedings of the 30th ACM International Conference on Architectural Support for Programming Languages and Operating Systems, Volume 1}{March 30-April 3, 2025}{Rotterdam, Netherlands}
\acmBooktitle{Proceedings of the 30th ACM International Conference on Architectural Support for Programming Languages and Operating Systems, Volume 1 (ASPLOS '25), March 30-April 3, 2025, Rotterdam, Netherlands}\acmDOI{10.1145/3669940.3707237}
\acmISBN{979-8-4007-0698-1/25/03}

\usepackage[]{hyperref}

\ccsdesc[500]{Computing methodologies~Distributed programming languages}
\keywords{Dynamic Analysis; Runtime Systems; Tracing}

\usepackage[nofillcomment,linesnumbered,ruled,vlined,noend]{algorithm2e} % added

\SetCommentSty{mycommfont} % added

\usepackage{amsfonts}
\usepackage{amsmath}
\usepackage{cleveref}
\usepackage{graphicx}
\usepackage{listings}
\usepackage{subcaption}
\usepackage{balance}

\usepackage[]{hyperref}
\theoremstyle{definition}

\definecolor{keywordcolor}{rgb}{0.5,0,0.5}
\definecolor{textgray}{gray}{0.4}
\definecolor{mygray}{rgb}{0.5,0.5,0.5}
\title{Automatic Tracing in Task-Based Runtime Systems}
\author{Rohan Yadav}
\affiliation{%
    \institution{Stanford University}
    \city{Stanford}
    \state{California}
    \country{USA}
}
\email{rohany@cs.stanford.edu}
\author{Michael Bauer}
\affiliation{%
    \institution{NVIDIA}
    \city{Santa Clara}
    \state{California}
    \country{USA}
}
\email{mbauer@nvidia.com}
\author{David Broman}
\affiliation{%
    \institution{KTH Royal Institute of Technology}
    \city{Stockholm}
    \country{Sweden}
}
\email{dbro@kth.se}
\author{Michael Garland}
\affiliation{%
    \institution{NVIDIA}
    \city{Santa Clara}
    \state{California}
    \country{USA}
}
\email{mgarland@nvidia.com}
\author{Alex Aiken}
\affiliation{%
    \institution{Stanford University}
    \city{Stanford}
    \state{California}
    \country{USA}
}
\email{aiken@cs.stanford.edu}
\author{Fredrik Kjolstad}
\affiliation{%
    \institution{Stanford University}
    \city{Stanford}
    \state{California}
    \country{USA}
}
\email{kjolstad@cs.stanford.edu}

\renewcommand{\shortauthors}{Rohan Yadav et al.}

\newcommand{\name}{Apophenia}
\newcommand{\cunumeric}{cuPyNumeric}
\newcommand{\numpy}{NumPy}

\definecolor{todocolor}{rgb}{0.8,0,0}
\definecolor{editcolor}{rgb}{0,0,0.8}
\newcommand{\TODO}[1]{{\color{todocolor}#1}}
\newcommand{\IGNORE}[1]{}

\begin{abstract}

Implicitly parallel task-based runtime systems often
perform dynamic analysis to discover dependencies in and
extract parallelism from sequential programs.
Dependence analysis becomes expensive as task granularity
drops below a threshold.
Tracing techniques have been developed where
programmers annotate repeated program fragments (traces)
issued by the application,
and the runtime system memoizes the dependence analysis
for those fragments, greatly reducing overhead when the
fragments are executed again.
However, manual trace annotation can be brittle and
not easily applicable to complex programs built through
the composition of independent components.
We introduce \name{}, a system that automatically
traces the dependence analysis of task-based runtime systems,
removing the burden of manual annotations from
programmers and enabling new and complex programs to be traced.
\name{} identifies traces dynamically through a series
of dynamic string analyses, which find repeated program
fragments in the stream of tasks issued to the runtime system.
We show that \name{} is able to come between 0.92x--1.03x the performance of manually traced programs,
and is able to effectively trace previously untraced programs to yield speedups of between 0.91x--2.82x on the Perlmutter and Eos supercomputers.

\end{abstract}

\makeatletter
\gdef\@copyrightpermission{
  \begin{minipage}{0.3\columnwidth}
   \href{https://creativecommons.org/licenses/by/4.0/}{\includegraphics[width=0.90\textwidth]{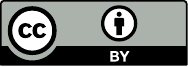}}
  \end{minipage}\hfill
  \begin{minipage}{0.7\columnwidth}
   \href{https://creativecommons.org/licenses/by/4.0/}{This work is licensed under a Creative Commons Attribution International 4.0 License.}
  \end{minipage}
  \vspace{5pt}
}
\makeatother

\begin{document}

\maketitle
% \pagestyle{plain} % should come right after \maketitle

% Note: 11 pages limit.

\section{Introduction}

Implicitly parallel programming systems~\cite{legion, ray, starpu, parsec, spark} automatically 
extract parallelism from a sequential source program through different forms
of dynamic dependence analysis.
Automatic parallelization and communication inference
has enabled composable high-level libraries~\cite{legate-numpy, legate-sparse} to
be built on top of implicitly parallel task-based runtime systems.
However, the cost of the dependence analysis affects the performance of 
implicitly parallel systems at scale and places a floor on the minimum 
problem size that can be executed efficiently~\cite{task-bench}.
Applications with tasks that are too small to amortize the cost of dependence
analysis is dominated by it and run at low efficiency.

To improve the performance of implicitly parallel task-based runtime systems,
researchers have proposed techniques~\cite{dynamic-tracing, omid-memoization} to memoize,
or \emph{trace}, the dependence analysis.
Tracing records the results of the dependence analysis for 
an issued program fragment, and then replays the results of the analysis the next time
an identical program fragment is issued.
Tracing has been shown to yield significant speedups by eliminating
the cost of the dependence analysis on iterative programs.
For example, tracing can reduce the per-task overhead in the Legion~\cite{legion} runtime system from $\sim$1ms to
$\sim$100$\mu$s~\cite{dcr}, widening the scope of 
applications for which task-based runtime systems
can be effective.

A significant limitation of existing tracing techniques is that
they require the programmer to annotate repeatedly issued program fragments
with stop/start markers for the runtime system.
Programmer inserted annotations derail an important feature of
implicitly parallel programming systems---their correctness under program composition.
As users develop modular programs that pass data from one component to another,
the runtime system ensures that computations launched by different modules maintain sequential semantics by implicitly inserting the necessary data movement and synchronization.
%are correctly synchronized against each other, and the necessary
%data is moved between the computations of different modules.
%
However, programmer introduced trace annotations do not obey these
composition principles, and the correct placement of 
trace annotations when composing complex software becomes unclear.
Functions defined in a third-party library may contain operations
that cannot by traced by a practical tracing implementation, or may
issue a different sequence of operations on each invocation.
Each of these cases result in runtime errors, due to the incorrect
trace annotations constructing an ill-formed sequence of operations.
Furthermore, even simple programs using high-level implicitly 
parallel libraries can have traces that do not correspond to syntactic
loop structures in the source program, making it difficult
to correctly place tracing annotations.
We elaborate on such an example program in \Cref{sec:motivating-example}.

% At a surface level, the current requirement of programmer annotations to use
% tracing techniques may seem marginal, but changing trends in
% the consumers of and the programs written in implicitly parallel task-based
% runtime systems exacerbate this restriction.
% %
% First, as high-level tasking libraries are built that target users who are not 
% parallel programming experts, each new ``magic'' annotation that must be added
% for performance reasons is a barrier to adoption and a seamless user experience.
% %
% As users develop modular programs that compose multiple libraries, the 
% correct placement of trace annotations also becomes unclear.
% %
% Functions defined in a third-party library may contain operations
% that, for a practical tracing implementation, cannot be traced or
% may issue a different sequence of operations on each call.
% %
% Finally, individual high-level libraries built with tasking runtimes may have
% unpredictable repetition patterns, as shown in \Cref{fig:legate-tracing-example}, where an annotation
% placed around a syntactic loop does not correspond to a repeated sequence
% of operations at runtime.

In order to improve programmer productivity and to enable the tracing of modular high-level
programs, implicitly parallel task-based systems should
automatically identify repeated sequences of operations, memoize their analysis results
and cheaply replay the analysis as needed.
We call this the problem of \emph{automatic trace identification}, which is similar to
Just-In-Time (JIT) compilation in the context of dynamic language implementations~\cite{self-jit, tracemonkeyjit, hotspotjit}.
JIT compilers for dynamic languages interpret bytecode during program startup, and compile
bytecode to native instructions as repeatedly invoked program fragments become hot.
Following this architecture, implicitly parallel task-based runtimes should
interpret issued operations with a dynamic dependence analysis,
and switch to an analysis-free compiled execution once repeated sequences of
operations are encountered.

% While inspired by JIT compilers, the application of these ideas to
% implicitly parallel task-based runtime systems results in a different set of challenges 
% --- the identification of program fragments to memoize becomes a key challenge.

We introduce our system \name{}\footnote{\name{} is the tendency to notice patterns between unrelated things.}, that acts as a JIT compiler
for the dependence analysis of an implicitly parallel task-based runtime system.
The key challenge that \name{} faces is the 
\emph{identification} of repeated sequences of operations produced by the target program.
Unlike JIT compilers, the input to a task-based runtime system is a stream
of tasks that lacks information about control flow such as basic block labels
or function definitions.
As such, \name{} cannot rely on these code landmarks
or predictable execution flow to identify repeated sequences of operations.
Instead, \name{} analyzes the input stream of operations to find
repetitions by solving a series of online string analysis problems.
%
% We define what are the desirable properties of traces to identify (\Cref{sec:good-traces}),
% and develop a series of algorithms to dynamically identify traces with
% these properties (\Cref{sec:trace-identification}).
%
% We show how this analysis can be done efficiently within
% the context of an asynchronous distributed runtime system.

To demonstrate \name{}, we develop an implementation within the Legion~\cite{legion} 
runtime system as a  front-end component that sits between the application and 
Legion's dependence analysis engine.
As operations are issued to Legion, \name{} performs a series of dynamic
analyses to identify repeatedly issued sequences of operations, and correspondingly
invokes Legion's tracing engine~\cite{dynamic-tracing} to memoize and replay
dependence analysis on these sequences.
While our prototype targets Legion, we believe that
the ideas in \name{} can be directly applied to
other task-based runtime systems that perform
a dynamic dependence analysis.

The specific contributions of this work are:
\begin{enumerate}
    \item A formulation of the desirable properties of traces to identify (\Cref{sec:good-traces}).
    \item Algorithms to dynamically identify traces in an application's stream of operations (\Cref{sec:trace-identification}).
    \item An implementation of \name{} that targets the Legion~\cite{legion} runtime system.
\end{enumerate}

To evaluate \name{}, we apply it to the largest and most complex
Legion applications written to this date, including production-grade
scientific simulations and machine learning applications.
We show that on up to 64 GPUs of the Perlmutter and Eos supercomputers,
\name{} is able to achieve between 0.92x--1.03x the performance
of manually traced code, and is able to effectively trace previously
untraced code built from the composition of high-level components to yield end-to-end speedups of between 0.91x--2.82x.
As such, \name{} is able to insulate programmers against
the overheads of task-based runtime systems on varying applications and problem sizes, transparently and without programmer intervention.

% \TODO{Contributions:

% \begin{itemize}
%     \item Definition / crystallization of intuition about what are good traces to find
%     \item A series of algorithms to identify good traces in a dynamic stream of operations
%     \item An implementation that is able to identify traces in complex, real-world programs.
% \end{itemize}

% }

\section{Motivating Example}\label{sec:motivating-example}

\begin{figure}
\begin{subfigure}[h]{0.49\linewidth}
\begin{center}
\begin{tabular}{c}
\begin{lstlisting}
import cupynumeric as np
# Generate random system.
A = np.random.rand(N,N)
b = np.random.rand(N)
# Initialize solution and
# extract diagonal.
x = np.zeros(A.shape[1])
d = np.diag(A)
R = A - np.diag(d)
# Jacobi iteration.
for i in range(iters):
  x = (b - np.dot(R, x)) / d
\end{lstlisting}
\end{tabular}
\end{center}
\caption{Python source code.}
\label{fig:legate-tracing-cunumeric}
\end{subfigure}\hfill
\begin{subfigure}[h]{0.49\linewidth}
\begin{center}
\begin{tabular}{c}
\begin{lstlisting}
DOT(R, ~x1~, t1)
SUB(b, t1, t2)
DIV(t2, d, ~x2~) # Iteration 1
DOT(R, ~x2~, t1)
SUB(b, t1, t2)
DIV(t2, d, ~x1~) # Iteration 2
DOT(R, ~x1~, t1)
SUB(b, t1, t2)
DIV(t2, d, ~x2~) # Iteration 3
DOT(R, ~x2~, t1)
SUB(b, t1, t2)
DIV(t2, d, ~x1~) # Iteration 4
\end{lstlisting}
\end{tabular}
\end{center}
\caption{Main loop task stream.}
\label{fig:legate-tracing-task-stream}
\end{subfigure}
\caption{A \cunumeric{}~\cite{legate-numpy} program and the stream of tasks it issues at runtime. An intuitive trace around the main loop does not correspond to a repeated program fragment.}
% \caption{A \cunumeric{}~\cite{legate-numpy} program demonstrating the necessity of automatic trace identification.}
\label{fig:legate-tracing-example}
\end{figure}

We now show an example of high-level implicitly
parallel code where it is difficult for a programmer to place tracing annotations.
As part of developing the example, we provide necessary background
on the Legion~\cite{legion} runtime system.

% A surprisingly difficult program to manually place 
% trace annotations in is shown in \Cref{fig:legate-tracing-example}.
%
\Cref{fig:legate-tracing-cunumeric} performs Jacobi iteration
using \cunumeric{}~\cite{legate-numpy}, a distributed drop-in
replacement for \numpy{}.
\cunumeric{} distributes \numpy{} through a dynamic translation
to Legion.
\cunumeric{} implements \numpy{} operations by issuing one or more 
Legion \emph{tasks}, which are designated functions registered 
with the runtime system.
Each \numpy{} array is mapped to a Legion \emph{region}, which is a
multi-dimensional array tracked by Legion.
Each task takes a list of regions as arguments.
The stream of tasks launched by the main loop of the \cunumeric{} program
is in \Cref{fig:legate-tracing-task-stream}.
For each task, the first two arguments denote the inputs, while the third
argument is the output.
%
% \TODO{
% We describe the label next to each task in \Cref{sec:good-traces}.
% }
%
Legion extracts parallelism from the issued stream of tasks by analyzing
the data dependencies between tasks and the usage of their region arguments~\cite{legion-dep-analysis}.

To trace a program fragment, the programmer issues
a \texttt{tbegin(id)} call (standing for ``trace begin'') before and a \texttt{tend(id)} call 
after the fragment.
The first time Legion executes a trace with a 
particular \texttt{id},
it records the results of the dependence analysis, and then
replays the results when executing the same trace \texttt{id} again~\cite{dynamic-tracing}.
For a trace to be valid, the sequence of tasks and their region arguments
encapsulated by \texttt{tbegin(id)} and \texttt{tend(id)} calls must
be exactly the same for a given \texttt{id}.
The same region arguments must be used across trace invocations as the
dependence analysis is affected by the usages of the regions and how they are partitioned.
While we consider regions for a Legion implementation of \name{},
this restriction generalizes to any form of argument that
affects the dependence analysis.

A natural attempt to trace the program in \Cref{fig:legate-tracing-cunumeric} would
place the \texttt{tbegin} and \texttt{tend} around the body
of the main \texttt{for} loop.
However, this annotation results in an invalid trace %(raising a runtime error), 
, for a subtle reason that requires knowledge of the
internals of \cunumeric{}.
The problem with this natural annotation is the loop-carried use of the Python
variable \texttt{x}, which is bound to different \cunumeric{} arrays (regions)
at different points of execution.
Upon entering loop iteration $i$, \texttt{x} is bound to a region arbitrarily 
named \texttt{x1}, which is used as an argument for the first \texttt{dot} operation.
As execution proceeds, \cunumeric{} allocates a new region \texttt{x2} for the result of the division
with \texttt{d}, and binds the Python variable \texttt{x} to the region \texttt{x2}.
Therefore, the next iteration $i+1$ issues a \texttt{dot} on
\texttt{x2}, causing iteration $i+1$ to issue a different sequence of tasks than iteration $i$!
Issuing a different sequence of tasks with the same trace \texttt{id}
is a violation of the conditions to use tracing, and the runtime system
may either raise an error or fall back to the expensive dependence analysis.
This program illustrates a real-world case where
abstraction and composition make it difficult to apply
the low-level tracing technique.

To correctly trace the program in \Cref{fig:legate-tracing-cunumeric}, a programmer
must either add trace annotations around every two iterations of the main loop,
or use two different trace ID's for each different iteration's repetition pattern.
This steady state of groups of two iterations is achieved because when \texttt{x} is assigned, the region it refers to can be collected and immediately reused by \cunumeric{}.
Relying on this steady state is brittle, as the addition of more operations in the main loop or
a change in \cunumeric{}'s region allocation policy could perturb the way in which
the necessary steady state for tracing is achieved.
Instead, \name{} dynamically analyzes the stream of tasks and automatically
discovers what fragments of the application should be traced, removing this
concern from the programmer.

% Background / example? (maybe, have to write it and see how it goes)

\section{What Are Good Traces?}\label{sec:good-traces}

The overarching goal of \name{} is to reduce the amount of time the runtime spends performing dynamic dependence analysis by selecting traces to replay.
A simple model of a tasking runtime system's dependence analysis is that the runtime spends time $\alpha$ analyzing each task.
The first time a trace is issued, the dependence analysis results are memoized,
so the runtime spends time $\alpha_m$ (memoization time) on each task in the trace, where $\alpha_m$ is slightly larger than $\alpha$.
Then, on subsequent executions of the trace, there is some constant $c$ amount of overhead to replaying the trace, but every task in the trace only incurs an analysis cost of $\alpha_r$ (replaying time), where $\alpha_r \ll \alpha$.

Using this model of the runtime system, we derive several properties of traces that \name{} should find.
First, the selected traces should maximize the number of traced operations to minimize the number of tasks that contribute an $\alpha$ to the overall analysis cost.
Next, the selected traces should be relatively long so that the constant replay cost $c$ does not accumulate.
Finally, the set of selected traces should be small, so that \name{} does not continually memoize new traces and pay $\alpha_m$ per task in each new trace.
Intuitively, the ideal set of traces corresponds to the loops in the target program.

\begin{figure}
    \centering
    \includegraphics[width=\linewidth]{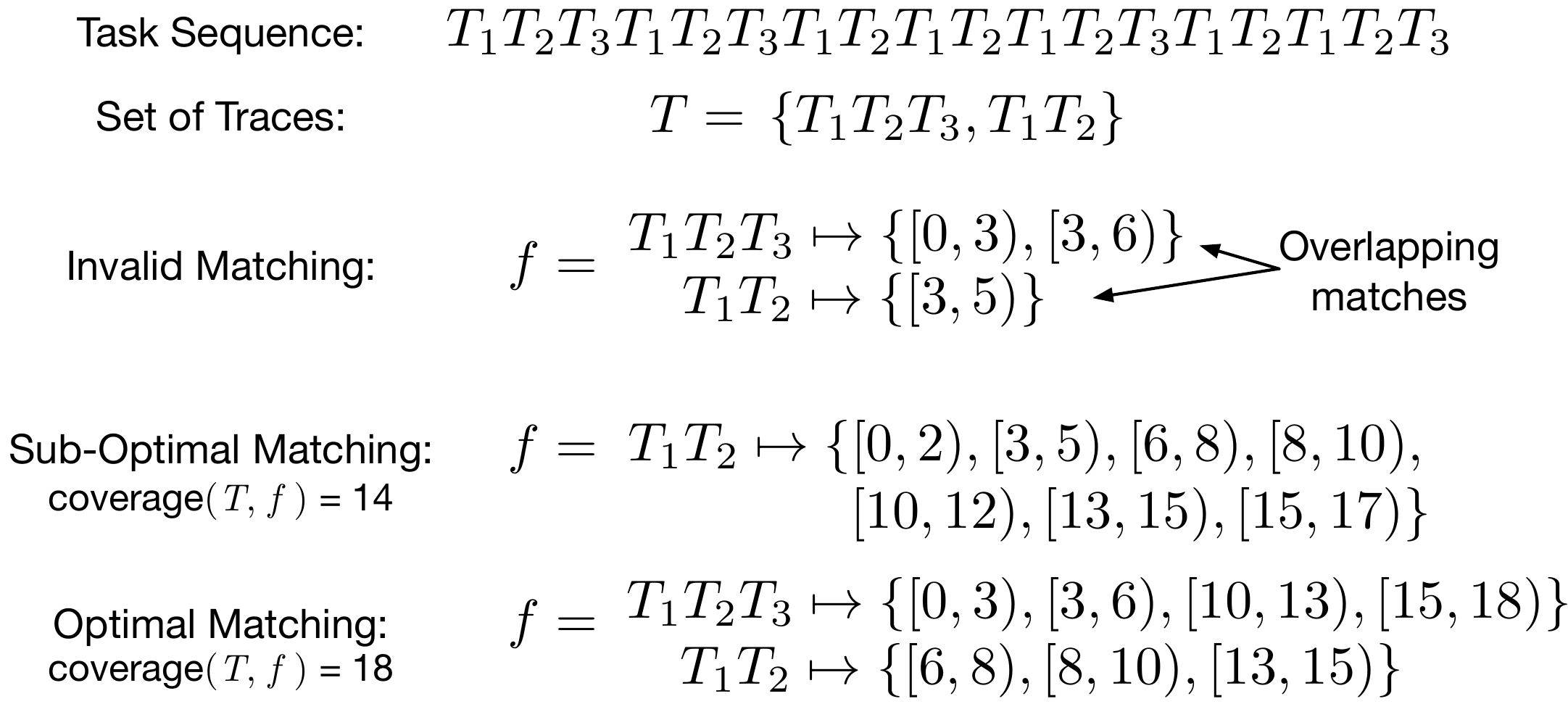}
    \caption{Example of a task stream and fixed trace set $T$ with an invalid matching function $f$, and two matching functions with different $\textsf{coverage}(T, f)$.}
    \label{fig:opti-problem-example}
\end{figure}

We now concretize the good traces that \name{} should find as the solutions of a concrete optimization problem.
Consider the sequence of tasks $S$ constructed from a complete execution of the target program.
A system for automatic trace identification must construct from $S$
\begin{itemize}
    \item A set of traces $T$, containing sub-strings of $S$,
    \item A function $f : T \rightarrow \textsf{interval set}$, mapping each $t \in T$ to a set of
    intervals in $S$ that are \emph{matched} by $t$,
\end{itemize}
that maximizes the \emph{coverage} of $f$, defined by
$\textsf{coverage}(T, f) = \sum_{t\in T} \sum_{i \in f(t)} |i|$,
subject to the
constraints
\begin{enumerate}
    \item $\forall t \in T$, $t$ is longer than a minimum length,
    \item $\bigcup_{t \in T} f(t)$ is a disjoint set of intervals.
\end{enumerate}
Multiple solutions exist for this problem, so we prefer solutions that
first maximize the number of matched intervals ($\sum_{t \in T} |f(t)|$),
and then minimize the total number of selected traces ($|T|$).
Maximizing $\textsf{coverage}(T, f)$ directly minimizes the number
of untraced tasks, and selecting a small set of traces that repeats many times
minimizes the memoization cost of $\alpha_m$ per task.
Finally, a minimum length is placed on traces to ensure that the constant replay
cost $c$ can be effectively amortized.
We present a concrete problem instance and example solutions in \Cref{fig:opti-problem-example}.

The presented optimization problem precisely defines the properties of traces
that a system like \name{} should attempt to find, but it does not directly
yield an algorithm to discover good solutions.
Additionally, the optimization problem is structured in a post-hoc formulation,
where an optimal solution is constructed from the results of the entire program execution.
In practice, a system like \name{} must construct the solution $(T, f)$ in an online manner,
using the currently visible prefix of the sequence of tasks launched by the application.
In the next section, we discuss algorithms for dynamically finding good solutions to this optimization problem through a set of string processing algorithms.

\section{Trace Identification}\label{sec:trace-identification}

\begin{figure}
    \centering
    \includegraphics[width=\linewidth]{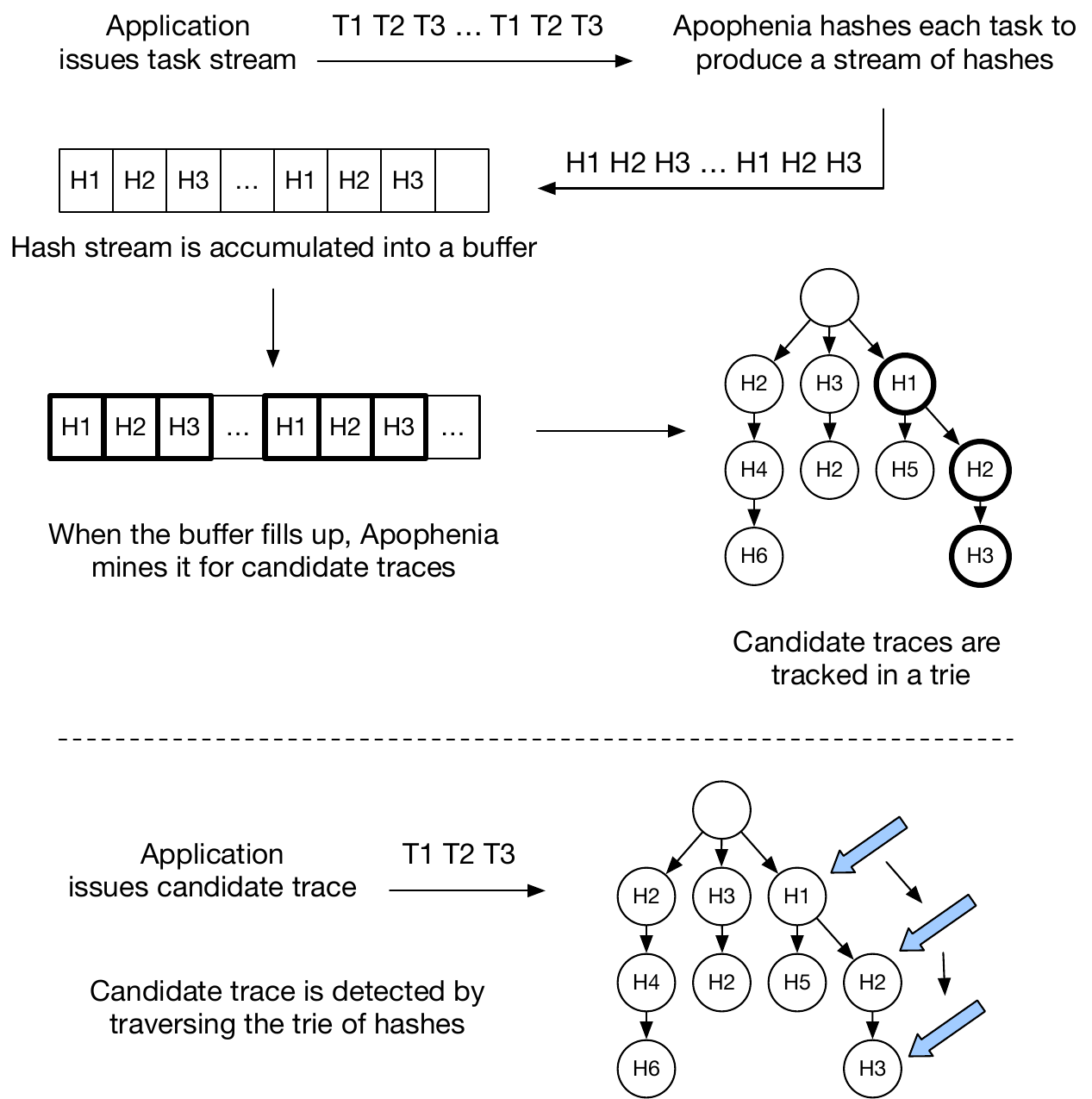}
    \caption{Visualization of \name{}'s dynamic analysis.}
    \label{fig:trace-lifetime}
\end{figure}

Dynamically finding good traces requires
processing information about the tasks seen so far, and then using
that information to record and replay traces in the future.
An overview of \name{}'s dynamic analysis procedure is sketched 
in \Cref{fig:name-overview}.
\name{} has two components that correspond
to the targets of the optimization problem 
in \Cref{sec:good-traces}.
The \emph{trace finder} constructs the candidate set of 
traces $T$ by
accumulating the tasks issued by the application 
into a buffer, and asynchronously mining the buffer to 
find candidate traces.
The \emph{trace replayer} then constructs the matching 
function $f$ by ingesting the candidate traces into a trie, 
and identifying candidate traces in the application stream 
by maintaining pointers into the trie that represent potential matches.
\name{} intercepts calls to target runtime's \texttt{ExecuteTask} function,
and forwards a potentially different set of tasks and trace markers
to the runtime.
A concrete example of how \name{} identifies a trace in an application
is shown in \Cref{fig:trace-lifetime}.
We now describe each of these components in detail.

\begin{algorithm}[t]
\caption{\name{}'s Dynamic Analysis.}
\label{fig:name-overview}
\footnotesize

\SetKwProg{ExecuteTask}{ExecuteTask}{}{}
\SetKwProg{TraceFinder}{TraceFinder}{}{}
\SetKwProg{TraceReplayer}{TraceReplayer}{}{}

\SetKwFunction{FExecuteTask}{ExecuteTask}
\SetKwFunction{FTraceFinder}{TraceFinder}
\SetKwFunction{FTraceReplayer}{TraceReplayer}
\SetKwFunction{FHash}{Hash}
\SetKwFunction{FShouldAnalyzeHistory}{ShouldAnalyzeHistory}
\SetKwFunction{FGetAnalysisSubset}{GetAnalysisSubset}
\SetKwFunction{FFindRepeats}{FindRepeats}
\SetKwFunction{FMaybeClearHistory}{MaybeClearHistory}
\SetKwFunction{FTrie}{Trie}
\SetKwFunction{FIngestCandidates}{IngestCandidates}
\SetKwFunction{FAdvanceActiveCandidates}{AdvanceActiveCandidates}
\SetKwFunction{FFilterInvalidCandidates}{FilterInvalidCandidates}
\SetKwFunction{FFilterCompletedCandidates}{FilterCompletedCandidates}
\SetKwFunction{FSelectReplayTrace}{SelectReplayTrace}
\SetKwFunction{FExecuteAndReplay}{ExecuteAndReplay}
\SetAlgoLined
\DontPrintSemicolon

\tcc{Initialize token history buffer $B$ and pending async analyses $J$.}
$B, J \gets [], []$\;

\tcc{Initialize trie of candidates $C$, potential current traces $A$, and pending tasks $P$.}
$C, A, P \gets \FTrie{}, [], []$ \;

% $B \gets []$ \;
% $J \gets []$ \;
% $B \gets []$ \tcp*{Token history buffer.}
% $J \gets []$ \tcp*{Pending asynchronous analyses.}
\tcc{Discussed in \Cref{sec:token-buffers}.}
\TraceFinder{$(H)$}{
    $B \gets B + [H]$\;
    \If{\FShouldAnalyzeHistory{$B$}}{
        \tcc{What subset of the history to analyze is discussed in \Cref{sec:ruler-func}.}
        $B' \gets $ \FGetAnalysisSubset{$B$}\;
        \tcc{Find repeated sub-strings.}
        $j \gets $ \textsf{async} \FFindRepeats{$B'$} \;
        $J \gets J + [j]$ \;
        $B \gets $ \FMaybeClearHistory{$B$}
    }
}

% $A \gets []$ \; 
% $P \gets []$ \; 
% $C \gets \FTrie{}$ \tcp*{Trie to track candidate traces.}
% $A \gets []$ \tcp*{List of potential trace matches.}
% $P \gets []$ \tcp*{List of pending tasks.}
\tcc{Discussed in \Cref{sec:replaying-traces}.}
\TraceReplayer{$(T, H)$} {
  \If{$\exists~j \in J, $ $j$ is complete}{
    \FIngestCandidates{$j, C$} \;
  }
  $P \gets P + [T]$\;
  \tcc{Advance all potential traces by $H$ in the trie if possible. Remove impossible traces, and extract fully matched candidates.}
  $A \gets $ \FAdvanceActiveCandidates{$C, A, H$}\;
  $A \gets $ \FFilterInvalidCandidates{$C, A$} \;
  $D, A \gets $ \FFilterCompletedCandidates{$C, A$} \;
  \If{$|D| > 0$} {
    \tcc{Select one of the pending candidates to replay. Execute any tasks before it, and issue a trace replay for the candidate.}
    $R \gets $ \FSelectReplayTrace{$D, P, A$} \;
    $P, A \gets$ \FExecuteAndReplay{$R, P, A$} \;
  }
}

\tcc{Applications issue tasks through \name{}'s ExecuteTask function.}
\ExecuteTask{$(T)$} {
    $H \gets $ \FHash{$T$}\;
    \FTraceFinder{$H$}\;
    \FTraceReplayer{$T$, $H$}\;
}
    
\end{algorithm}

\subsection{A Stream of Tokens}\label{sec:hashing-tasks}

An insight of our work is that automatic trace identification
is inherently an online string analysis problem of finding repeated sub-sequences
in the application's task stream.
As seen in \Cref{fig:legate-tracing-task-stream}, the task stream is not just
a list of identifiers---tasks have arguments that must also be the same across
iterations to be used in a trace.
To capture all aspects of a task that can affect the dependence analysis,
\name{} constructs a hash from each task and its region arguments.
Converting the input stream of tasks into a stream of hash tokens enables
more direct application of string processing techniques, and straightforward
handling of traceable operations that are not tasks. %, such as copies and fills.

\subsection{Finding Traces With High Coverage}\label{sec:token-buffers}

\name{}'s trace finder records tasks as they are issued 
by the application into a buffer (we describe a refinement to this scheme in \Cref{sec:ruler-func}).
Once the buffer fills up, \name{} launches an asynchronous analysis of the buffer to
find a set of traces within the buffer that maximize the coverage of
the buffer.
We discuss previous ideas that are related to this goal, and then describe the
solution used in \name{}.\footnote{We discuss more related work in \Cref{sec:related-work}.}
% We first survey related techniques for such an analysis, and then describe the solution
% used in \name{}.

\paragraph{Existing Techniques}

% While useful for applications like plagiarism detection,
% we found that these techniques are not directly applicable to
% the problem of selecting repeated sub-strings within an
% individual document.

The Lempel-Ziv family of algorithms use repeated sub-strings for compression.
Algorithms like LZ77~\cite{LZ77, LZ78, LZSS} maintain a sliding window of previous tokens to search for repeats in when encoding upcoming tokens.
The LZW~\cite{LZW} algorithm avoids the use of a sliding window by only increasing the length of any candidate repeat by a single token at a time.
While not directly finding a set of repeats with high coverage,
similar algorithms that use a sliding window would need to maintain and search in a window the size of the analyzed buffer, resulting in quadratic time complexity.
In order to recognize a trace of length $n$, an LZW-style algorithm would also need to encounter the trace $n-1$ times.
We wanted an algorithm that is sub-quadratic in order to scale to large buffer sizes.
Real-world applications we discuss in \Cref{sec:evaluation} have traces that contain more than 2000 tasks, requiring token buffers of at least
twice that size to detect a single repeat.

Within the programming languages community, recent work by Sisco et al.~\cite{loop-rerolling}
used a technique called \emph{tandem repeat analysis}~\cite{tandem-repeats}
to find loops in the netlists that result from compiling hardware description languages.
A tandem repeat is a sub-string $\alpha$ that repeats contiguously within a larger
string $S$, such that $\alpha^k$ is a sub-string of $S$, for some $k$.
Despite the success that Sisco et al. had using tandem repeat analysis, we found that
even simple real world \cunumeric{} programs did not contain enough tandem repeats
for the analysis to reliably identify a trace set with high coverage.
The reason is that while these real-world programs tended to have repetitive main loops, there would
often be irregularly appearing computations such as convergence checks or statistics calculations that
occur infrequently between loop iterations.
As such, the strings that represented these programs would not contain tandem
repeats, but instead repeated sub-strings separated by other tokens.

A relaxation of tandem repeat analysis is to search for non-overlapping repeated sub-strings,
which removes the contiguity requirement on the repeats.
Concretely, given the string $ababab$, $abab$ is an overlapping repeat, while $ab$ is non-overlapping.
We could use non-overlapping repeated sub-strings to assemble
a set of traces $T$ and a disjoint mapping $f$ that achieves high coverage.
While there exist standard suffix-tree algorithms to find repeated sub-strings, we found that
the natural extensions of these algorithms to detect non-overlapping repeated sub-strings also resulted
in quadratic runtime complexity.

% , which we believed would not scale as the buffer sizes increased.
% %
% Buffer sizes must be large enough to contain multiple occurrences
% of traces found in real programs.
% %
% Real-world applications we discuss in \Cref{sec:evaluation} have traces that contain more than 2000 tasks, requiring token buffers of at least
% twice that size to have even a chance of detecting a single repeat.

\paragraph{Our Algorithm}
In this work, we design a repeat finding algorithm that is directly aware of the
optimization problem in \Cref{sec:good-traces} and runs in $O(n\log(n))$, where $n$
is the size of the token history buffer.
At a high level, our algorithm makes a pass through
a suffix array constructed from the input buffer to collect
a set of candidate repeats.
It then greedily selects the largest repeated sub-strings that do not
overlap with any previously chosen sub-strings.
Psuedocode for our algorithm is in \Cref{fig:repeats-algorithm}\footnote{We also present a standalone implementation of the algorithm available at \url{https://github.com/david-broman/matching-substrings}.}, which takes a string $S$ and returns a set of
sub-strings that achieve high coverage of $S$.
We assume that the reader is knowledgeable
about suffix arrays and their structural properties. However,
understanding the algorithm in \Cref{fig:repeats-algorithm} is not required to understand its usage in \name{}, as discussed in \Cref{sec:replaying-traces} and \Cref{sec:ruler-func}.

\begin{algorithm}[t]
\caption{Non-overlapping repeated sub-strings.}
\label{fig:repeats-algorithm}
\footnotesize
\SetKwProg{FindRepeats}{FindRepeats}{}{}

\SetKwFunction{FSuffixArray}{SuffixArray}
\SetKwFunction{FSort}{Sort}
\SetKwFunction{FDeduplicate}{Deduplicate}
\SetAlgoLined
\DontPrintSemicolon

\FindRepeats{$(S)$}{
    $SA, LCP \gets$ \FSuffixArray{$S$} \;
    \tcc{Candidates are tuples of string length, the repeated sub-string, and starting position.}
    $C \gets []$\;
    \ForEach{$i \in [0, |SA| - 1)$}{
        \tcc{Extract adjacent suffix array entries and their overlap length.}
        $s1, s2, p \gets SA[i], SA[i+1], LCP[i]$ \;
        \If{$[s1:s1+p) \cap [s2:s2+p) = \emptyset$}{
            \tcc{$S[s1:s1+p]$ and $S[s2:s2+p]$ are repeated strings that do not overlap in $S$, so they are candidates.}
            $r \gets S[s1:s1+p]$ \;
            $C \gets C + [(p, r, s1), (p, r, s2)]$ \;
        }
        \Else{
            \tcc{$S[s1:s1+p]$ and $S[s2:s2+p]$ overlap in $S$. Assume $s2 > s1$, the other case is symmetric. In this case, the overlap is a collection of repeats of $S[s1:s1+d]$, by the structure of the suffix array.}
            $d \gets s2 - s1$\;
            \tcc{Break prefix into two chunks of repeated pieces of $S[s1:s1+d]$.}
            $l \gets (p + d) / 2$ \;
            \tcc{Remove trailing tokens.}
            $l \gets l - (l \% d)$\;
            $r \gets S[s1:s1+l]$ \;
            $C \gets C + [(l, r, s1), (l, r, s1 + l)]$\;
        }
    }
    \tcc{Sort the candidates by decreasing length and by increasing sub-string and start position.}
    \FSort{$C$}\;
    \tcc{Greedily collect sub-strings that do not overlap with previously chosen sub-strings.}
    $I, R \gets [], []$ \;
    \ForEach{$(l, \_, s) \in C$} {
        \If{$[s, s+l)$ \text{does not intersect} $I$}{
            $I \gets I + [[s, s+l)]$\;
            $R \gets R + [S[s:s+l]]$\;
        }
    }
    \Return{$R$}
}
\end{algorithm}

% \begin{algorithm}
% \begin{algorithmic}
% \Procedure{FindRepeats}{$S$}:
%   \State $SA, LCP$ $\gets$ \Call{SuffixArray}{$S$}
% \EndProcedure
% \end{algorithmic}
% \end{algorithm}

\begin{figure}
    \centering
    \includegraphics[width=0.8\linewidth]{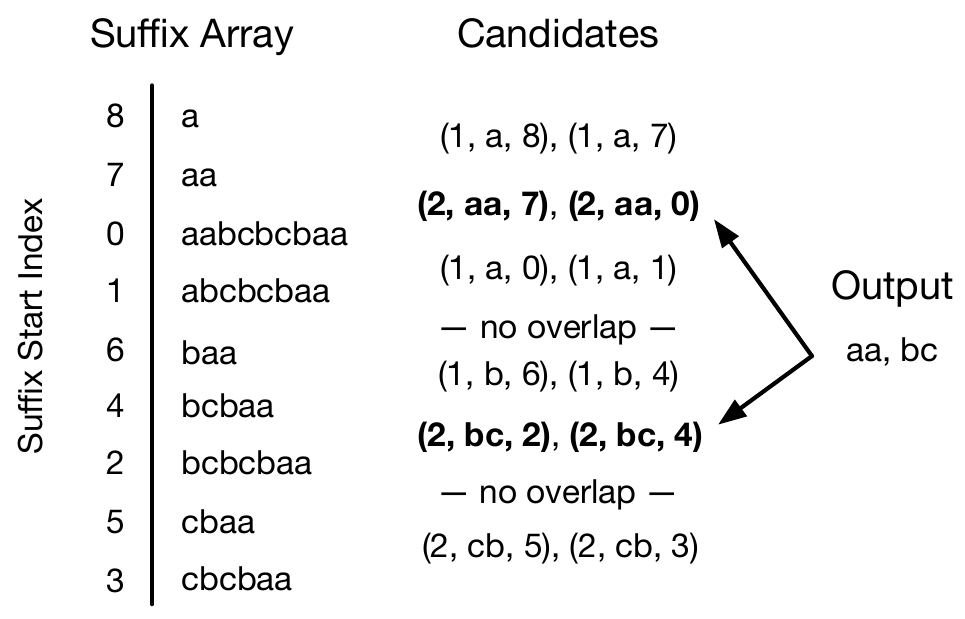}
    \caption{Execution of \Cref{fig:repeats-algorithm} on ``aabcbcbaa''. The candidates for each suffix pair is shown between the pair.}
    \label{fig:alg-example}
\end{figure}

As a first step, we construct a suffix array and longest
common prefix array from the input buffer of tokens.
We then iterate through adjacent pairs of suffixes to construct
a set of \emph{candidate repeats}, which are tuples of sub-strings
defined by their length, the repeated sub-string, and its starting position in $S$.
These candidates are constructed based on whether or not
the shared prefix between adjacent suffix array entries overlap.
Once all of the candidates have been constructed, we sort
the candidates to greedily select candidates in order of length, 
and select as many occurrences of a particular sub-string as possible.
We only select candidates that do not overlap with any previously
selected candidates, and then deduplicate the chosen set of
candidates as the result.
A sample execution of \Cref{fig:repeats-algorithm} is shown in \Cref{fig:alg-example}.

Our algorithm can be implemented with time complexity $O(n\log(n))$.
Linear time algorithms exist for suffix array and LCP array construction~\cite{kasai-alg}.
Two candidates are generated for each entry in the suffix array, so
sorting the candidates takes $O(n\log(n))$ time.
The interval intersection step can be reduced
to constant time by leveraging the candidate iteration order,
so the entire loop executes in $O(n)$ time.
In particular, an array of length $|S|$ can be maintained, and as
each candidate is selected, all positions covered by the candidate
are marked.
Then, as candidates are iterated over in decreasing length 
and increasing
start position order, interval intersection can be checked by checking
if the start or end of an interval is marked.
Finally, the deduplication can be done by generating a unique ID
for each candidate sub-string in the candidate generation phase,
and adjusting the candidate representation to be a tuple of
length, ID and starting position; using this sort order allows
deduplication to be done at each iteration of the candidate selection loop.

Our algorithm aims to find good solutions to the optimization
problem in \Cref{sec:good-traces} by identifying
long repeated sub-strings and selecting as many as possible
that do not overlap with each other.
We trade off between an optimal solution to the optimization
problem to instead find good solutions and maintain a lower
asymptotic runtime.
There are two such heuristics in our algorithm.
First, when adjacent suffix array entries have
a repetition, we consider only the maximal length repetition
instead of all sub-strings of the repetition.
Second, when we select which candidates to keep, we greedily
choose the largest candidates instead of performing
a bin-packing computation.
Our algorithm is guaranteed to find the longest repeated
sub-string, but due to the second heuristic, we cannot
provide theoretical guarantees about the other chosen sub-strings.
We show in \Cref{sec:evaluation} that \name{} using
our algorithm is able to identify good traces in complex, real-world
applications.

\subsection{Recognizing and Replaying Candidate Traces}\label{sec:replaying-traces}

\name{}'s trace replayer uses \Cref{fig:repeats-algorithm} to find 
candidate traces from the application's history of tasks.
In this section, we discuss how \name{}'s trace replayer 
identifies and selects these candidate traces from the
task stream to record and replay.
Our design of the trace replayer has two major goals.
First, the per-task overhead must be low,
as it is imperative for performance for the application to issue as many tasks into the
runtime as possible so that the runtime can either replay traces or perform dependence
analysis ahead of execution.
Slowing down the task launch rate would result in
exposed latency from various sources in the runtime.
Second, \name{} must balance exploration and exploitation when
selecting traces.
As more information about the application is gained, \name{} should switch
to better traces as it finds them.
However, \name{} should not leave a steady state of replaying a particular
trace until it is confident that performance can be improved, 
as memoization of the dependence analysis for new traces has a cost.

As discussed previously, \name{} accumulates a history of
tasks launched by the application and asynchronously uses \Cref{fig:repeats-algorithm} 
to select candidate traces.
Asynchronous analysis of task histories is important to avoid stalling
the application by waiting for the analysis to finish before accepting
the next task from the application.

When an asynchronous analysis completes, \name{} ingests the results into a
trie that maintains the current set of candidate traces.
Along with this trie, \name{} maintains a set of pointers into the trie that
represent potential matched traces.
As tasks are issued, \name{} updates the set of pointers by creating new pointers
for each new task, stepping any existing pointers down the trie if possible,
and removing any pointers that are made invalid.
Once a pointer reaches a leaf of the trie and has matched a trace, \name{} has
the option to forward the trace to the tasking runtime, wrapped by \texttt{tbegin} and \texttt{tend} calls.

\name{} uses a scoring function to select which matched trace to replay when
faced with multiple valid choices.
The scoring function is based on the length of the candidate trace multiplied
by a count of the number of times the trace has appeared.
In calculation of the score, we impose a maximum value of the count that
can be used, and exponentially decay the value of the count by how many tasks
have been encountered since the trace last appeared.
Finally, we increase the score slightly if a trace has already been replayed.

Our scoring function encodes heuristics about trace selection and aims to balance
exploration and exploitation.
We naturally prefer long traces over shorter ones, as longer traces have
the potential to eliminate more runtime overhead.
The capping of the appearance count allows for \name{} to eventually switch
from a trace that appeared early during program execution to a better trace
that appears later in the execution.
Next, decaying the appearance count ensures that a seemingly promising trace that
occurs infrequently, does not eventually hit a threshold, and disrupts a steady state.
Finally, since recording new traces has a cost, when faced with traces of a similar score,
we bias \name{} towards a trace it has already replayed.

\subsection{Achieving Responsiveness and Quality}\label{sec:ruler-func}

\name{}'s trace finder accumulates tasks into a buffer
and mines the buffer for traces using \Cref{fig:repeats-algorithm}.
An important question is what should the size of that buffer be?
The size of this buffer trades off between responsiveness of the \name{}'s
trace identification and the quality of traces \name{} is able to find.
With a small buffer, \name{} can identify traces early but will not be
able to identify traces in programs with large loops.
Meanwhile, a large buffer allows \name{} to identify long traces in 
complex applications but introduces significant startup delay in
smaller applications.

\begin{figure}
    \centering
    \includegraphics[width=0.6\linewidth]{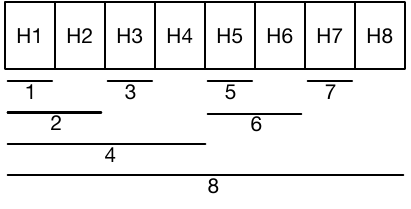}
    \caption{Visualization of \name{}'s buffer sampling strategy on a buffer of size 8. After processing the $i$'th task, \name{} mines the buffer slice labeled $i$.}
    \label{fig:ruler-func}
\end{figure}

We did not want end users to be required to 
continually adjust the buffer size parameter as their application changes.
As such, some strategy to adapt the buffer size along this tradeoff space is necessary.
We found that a strategy that attempts to dynamically resize the buffer based on 
what traces to find is unsatisfactory, as the system is unable to differentiate between
an application currently not repeating operations versus an application repeating a
sequence of operations larger than the buffer size.
Instead, we propose a strategy that selects a large fixed buffer size upfront, and then
samples smaller pieces of the buffer in a principled manner to be responsive to
the occurrence of short traces.

\name{} samples from
the buffer guided by the \emph{ruler function} sequence~\cite{ruler-function}, which
provides a practically useful sampling strategy with provable guarantees.
The ruler function counts the number of times a number can be evenly divided by two.
Applying it to the sequence $1, 2, 3, 4, \ldots$ yields the sequence $0, 1, 0, 2, \ldots$.
Raising the sequence to the power two yields $1, 2, 1, 4, \ldots$, which
we can interpret as subsets of the buffer to analyze.
For example, with a buffer size of four, as tasks arrive \name{} would first analyze the first task,
then the first two tasks, then the third task, and finally all four tasks.
A visualization of this sampling policy is in \Cref{fig:ruler-func}.
This sampling policy lets \name{} quickly react
to changes in the application by analyzing recent pieces of the buffer while
allowing larger traces to be found by infrequently analyzing longer components
of the buffer.
For example, sampling the full buffer in \Cref{fig:ruler-func} is required to
find a trace that repeats in positions H2-H4 and H5-H7.
In practice, we use the exponentiated ruler function as the multiples of a larger constant
(such as 250) to sample the buffer with.
Finally, given that our algorithm in \Cref{sec:trace-identification} runs in $O(n\log(n))$,
we show that our sampling strategy increases the total runtime complexity of
processing the buffer by only an extra $\log$ factor, yielding a total of $O(n\log^2(n))$.
This technique enables all of the experiments in \Cref{sec:evaluation} to be run
with the same buffer size configuration parameter.

\section{Implementation Discussion}

We now discuss important aspects of a realistic 
implementation of \name{}.
In particular, we discuss the specifics of implementing \name{} in a
distributed context and a decision not to perform speculation when replaying traces.

\IGNORE{
In particular, we discuss the specifics of implementing \name{} in a
distributed context, a decision  not to perform speculation when replaying traces, and
the necessary extensions Legion's underlying tracing engine.
}

\subsection{Distributing the Analysis}\label{sec:distribution}

\name{}'s analysis as presented in \Cref{sec:trace-identification} is sequential,
processing tasks as they are issued by the application.
In a distributed setting, \name{} leverages Legion's \emph{dynamic control replication}~\cite{dcr} to act as a sequential analysis, except
for one component, which we discuss next.
With control replication, the application executes on each node
and Legion shards the dependence analysis and execution across nodes.
The main restriction of control replication is that the
application must issue the same sequence of tasks on every node.
We implement \name{} as a layer between the application and Legion,
meaning that \name{} intercepts calls into the Legion runtime from
the application and forwards a (possibly different) set of calls into Legion.
As such, \name{} inherits the control replication requirements of the application.
In particular, each node must agree on which
traces to replay and when during program execution to record
and replay the traces.

The only source of non-determinism in \name{} that may result
in divergent decisions between nodes is the asynchronous
processing of token buffers described in \Cref{sec:token-buffers}.
An instance of \name{} exists on each node of the target machine,
and each instance maintains a local history buffer of tasks
to run asynchronous analyses on.
The asynchronous analysis may complete earlier on one node
than another, resulting in that node replaying a trace
before another node has identified that trace as a candidate.
However, making the analysis synchronous would result
in stalling the application until analyses complete.
We resolve this tension by having each node agree on a count
of processed operations to issue before ingesting the
results of an asynchronous analysis.
If any node had to wait on an asynchronous analysis to complete,
all nodes increase their count of operations to wait on for the
next analysis.
This strategy reaches a steady state where analysis results are
ingested in a deterministic manner without stalling the application.

\subsection{(The Lack of) Speculation}

Speculation is a common technique in computer architecture to efficiently
execute programs with data-dependent control flow.
As \name{} has similarities to speculative components in architecture like
trace caches (\Cref{sec:related-work}), a natural
design decision was if \name{} should speculate
on whether traces would be issued by the application.
%
% % Concretely, if \name{} decided to replay the trace $T_1T_2T_3T_1T_2T_3$ and has seen the tasks
% % $T_1T_2T_3$ so far, should it wait until the second $T_1T_2T_3$ has been issued, or speculatively
% % issue the trace $T_1T_2T_3T_1T_2T_3$ and roll back if the next three issued tasks differ from
% $T_1T_2T_3$?
%
Our implementation of \name{} does not speculate and waits
for the entirety of a trace to arrive before issuing the trace to Legion.
The relative costs of different operations within the Legion
runtime system made the potential upside of speculation not
worth the implementation complexity.

Legion employs a pipelined architecture where a task
flows  through three stages: 1) the application phase,
where the task is launched (into \name{}), 2) the analysis phase, where
the task is analyzed or replayed as part of a trace, and
3) the execution phase, where the task is executed.
Depending on the cost ratio of the application and analysis
phases, speculation may be beneficial as \name{} waits
for an entire trace to pass through the application phase.
Legion's analysis phase is an order of magnitude
more expensive than the application phase, letting
the application phase run far ahead of the analysis phase.
Thus, waiting for an entire trace to be issued by the application
rarely stalls the pipeline and gets exposed in the overall runtime.
Thus, we found that designing a trace prediction
algorithm and implementing a backup-rollback-recover scheme
on speculation failures was not worth the complexity.

\section{Evaluation}\label{sec:evaluation}

\begin{figure*}
\begin{subfigure}[b]{0.49\textwidth}
\centering
\includegraphics[width=0.9\textwidth]{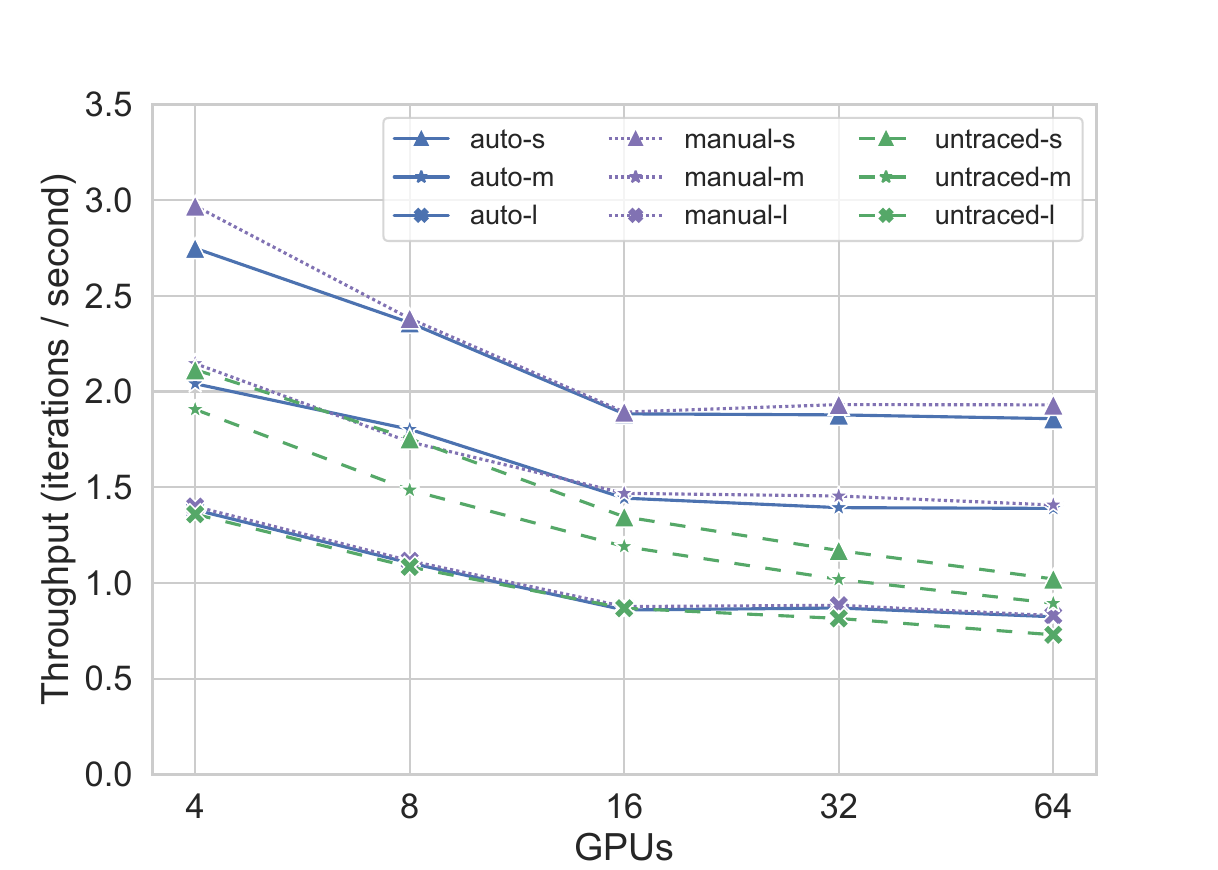}
\caption{S3D (Perlmutter)}
\label{fig:s3d-weak-scaling}
\end{subfigure}\hfill
\begin{subfigure}[b]{0.49\textwidth}
\centering
\includegraphics[width=0.9\textwidth]{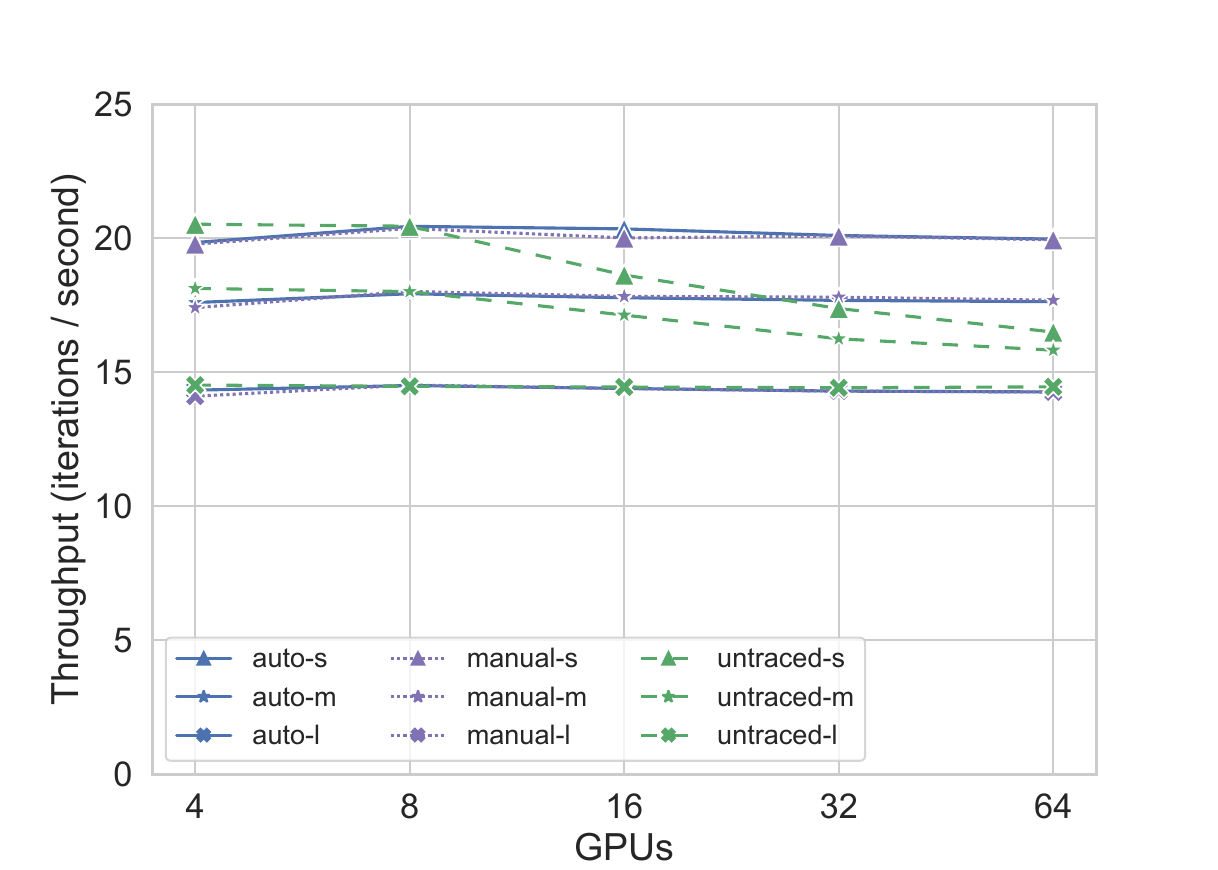}
\caption{HTR-Solver (Perlmutter)}
\label{fig:htr-weak-scaling}
\end{subfigure}
\caption{Weak scaling on previously traced Legion applications, where \name{} (``auto'') performs competitively.}
\label{fig:weak-scaling-perlmutter}
\end{figure*}

\begin{figure*}
\begin{subfigure}[b]{0.49\textwidth}
\centering
\includegraphics[width=0.9\textwidth]{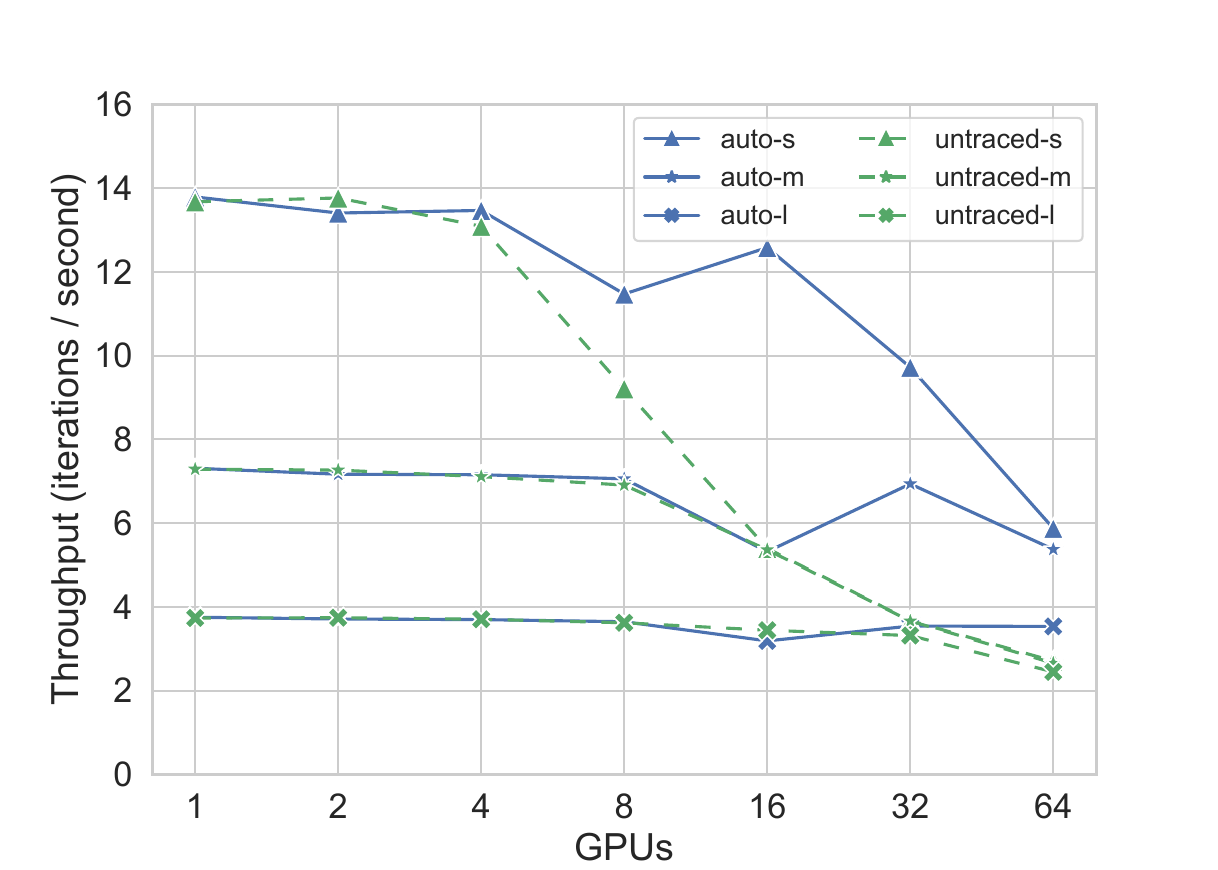}
\caption{CFD (Eos)}
\label{fig:cfd-weak-scaling}
\end{subfigure}\hfill
\begin{subfigure}[b]{0.49\textwidth}
\centering
\includegraphics[width=0.9\textwidth]{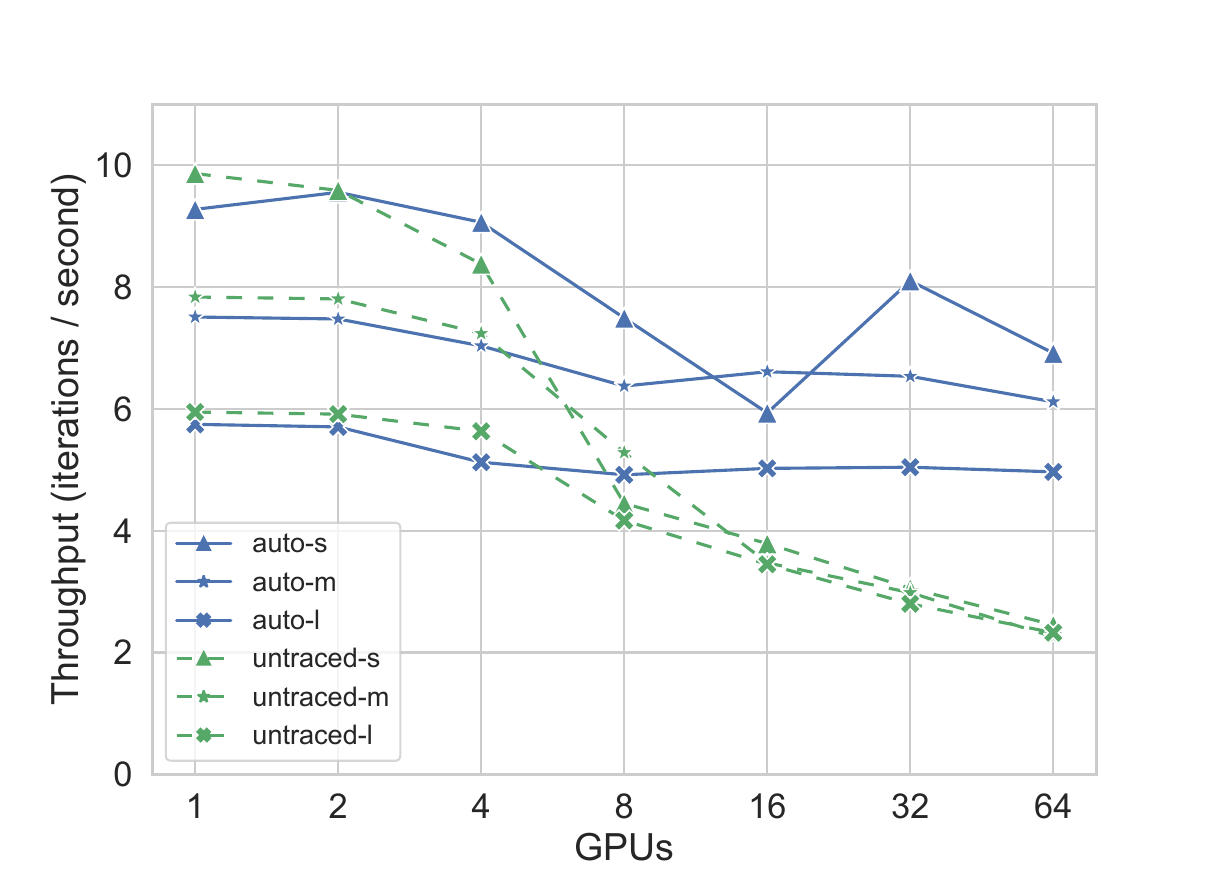}
\caption{TorchSWE (Eos)}
\label{fig:torchswe-weak-scaling}
\end{subfigure}
\caption{Weak scaling on
\cunumeric{} applications, where \name{} (``auto'') outperforms the untraced version.}
\label{fig:weak-scaling-eos}
\end{figure*}

\paragraph{Overview}
We evaluate \name{} on the largest and most complex Legion applications written
to date, including production scientific simulations and a distributed deep 
learning framework.
%
% We perform a mixture of weak-scaling (\Cref{sec:weak-scaling}) and strong-scaling
% (\Cref{sec:strong-scaling}) experiments and show that \name{} is able to match
% the performance of manually traced code when trace annotations already exist,
% and that \name{} can identify traces in programs that were previously not traced.
% %
% We then evaluate the overhead that \name{} imposes on Legion applications (\Cref{sec:task-overhead}) and visualize \name{}'s search process (\Cref{sec:search-visualization}).
%
Our results show that \name{} is able to effectively find traces in complex programs
with lower overhead, enabling programmers to experience the benefits of tracing
without manual effort and allowing a more general set of applications to be traced.

\paragraph{Experimental Setup}
We evaluated \name{} on the Eos and Perlmutter supercomputers.
Each node of Eos is an NVIDIA DGX H100, containing 8 H100 GPUs with
80 GB of memory and a 112 core Intel Xeon Platinum.
Each node of Perlmutter contains 4 NVIDIA A100 GPUs with 40 GB of memory
and a 64 core AMD EPYC 7763.
Nodes of Eos are connected with an Infiniband interconnect, while Perlmutter
uses a Slingshot interconnect.
We compile Legion on Eos with the UCX networking module, and use the GASNet-EX~\cite{gasnet-lcpc18}
networking module on Perlmutter.
We do not execute each application on both Perlmutter and Eos due to differences between the local environments on each machine.
In our experiments, we evaluate the relative performance differences between traced and untraced programs, and comparisons between machines are not significant.

\subsection{Weak Scaling}\label{sec:weak-scaling}

In this section, we discuss weak scaling results of applications using \name{},
as shown in \Cref{fig:weak-scaling-perlmutter} and \Cref{fig:weak-scaling-eos}.
In a weak scaling study, we increase the problem 
size as the size of the target machine grows to
keep the problem size per processor constant.
For each application, we perform a sweep over different sizes of the problem
to vary the task granularity, thus affecting the impact of runtime overhead.
These different problem sizes are denoted in the graph by the ``-s'', ``-m'' and ``-l''
label suffixes which stand for small, medium and large.
At smaller problem sizes, more runtime overhead can be exposed, while larger problem
sizes make it easier to hide runtime overhead.
In each weak-scaling plot, we report the steady-state throughput of each configuration
and problem size after a number of warmup iterations (discussed in \Cref{sec:task-overhead}).
We report throughput in iterations per second achieved by each configuration, so within
a particular problem size, higher is better; across problem sizes, the smaller problem
sizes will achieve a higher iterations per second than the larger problem sizes.

\paragraph{S3D}
S3D~\cite{s3d} is a production combustion chemistry simulation code that
has been developed over the course of many years by different scientists and engineers.
The Legion port of S3D implements the right-hand-side function of the Runge-Kutta
scheme, and interoperates with the legacy Fortran+MPI driver of the simulation.
The integration between Legion and the legacy Fortran+MPI code leads to various
constraints that the manual trace annotations interact with.
For example, during the first 10 iterations, a hand-off between Legion and Fortran+MPI 
must occur every iteration, while after the first 10 iterations a hand-off is required
only every 10 iterations.
While not unmanageable, these interactions have led to relatively complicated logic
to manually trace the main loop.
We scale S3D on Perlmutter, and compare the performance of \name{} to manually
traced and untraced versions of S3D.
The results are shown in \Cref{fig:s3d-weak-scaling}.
Even on a single node, tracing has a noticeable performance
impact on the smaller problem sizes and affects the scalability of
S3D.
\name{} achieves within 0.92x--1.03x of the performance of the manually
traced version, and between 0.98x--1.82x speedups over the untraced version.
Manual annotations can slightly outperform \name{} by leveraging
application knowledge to select traces that have lower replay overhead.

\paragraph{HTR}
HTR~\cite{htr} is a production hypersonic aerothermodynamics application.
HTR performs multi-physics simulations of hypersonic flows at 
high enthalpies and Mach numbers, such as for simulations of the 
reentry of spacecraft into the atmosphere.
Like S3D, we evaluate \name{}'s performance on HTR on Perlmutter, and
compare it against a manually traced version and an untraced version.
While HTR without tracing performs competitively to 
the traced version at small GPU counts, \Cref{fig:htr-weak-scaling} shows that tracing is
necessary for performance at scale.
\name{} achieves within 0.99x--1.01x of the performance of the manually
traced version, and between 0.96x--1.21x speedups over the untraced version.
%
% Interestingly, we see that on the largest problem size, the untraced version
% of HTR is slightly faster than with \name{} or the manually traced version.
% %
% \TODO{rohany: What's a reasonable explanation here that doesn't get into too many details...}

\paragraph{CFD}
CFD is a \cunumeric{} application that solves the Navier-Stokes equations
for 2D channel flow~\cite{cfd}.
Unlike S3D and HTR, there is not a manually traced version of 
CFD, due to the difficulties around composition discussed in \Cref{sec:motivating-example}.
Developing a manually traced implementation of CFD would require either rewriting
the application to remove any dynamic region allocation, or manual examination
of allocator logs to find the number of iterations in the steady state.
As a result, we compare CFD with \name{} to the standard untraced version on
different problem sizes, which is the performance that \cunumeric{} users are
able to achieve today.

\Cref{fig:cfd-weak-scaling} shows weak scaling results for CFD on Eos.
These results are similar to HTR, where leveraging tracing is necessary 
for performance at scale.
On the smallest problem size, even though the tracing removes a large amount of
runtime overhead, the tasks are too small to hide the communication latency
at larger scales, leading to the observed fall off in performance.
On larger problems, CFD with \name{} is able to maintain high performance while
the untraced version falls off, yielding between 
0.92x--2.64x speedups.

\paragraph{TorchSWE}
TorchSWE is a \cunumeric{} port of the MPI-based TorchSWE~\cite{torchswe}
shallow-water equation solver, and is the largest \cunumeric{} application
developed so far.
Similarly to CFD, there is no manually traced version to compare to.
However, unlike CFD, performing a rewrite of TorchSWE to enable manual
tracing would be difficult, as TorchSWE contains an order of magnitude
more lines of code.
Weak scaling results for TorchSWE on Eos are shown in \Cref{fig:torchswe-weak-scaling},
which show that TorchSWE is significantly impacted by Legion runtime overhead
without tracing.

These results demonstrate that there does not exist a problem size for
TorchSWE on Eos that can hide Legion's runtime overhead
without tracing.
Even the large problem size, which nearly reaches the GPU's memory
capacity, exposes Legion runtime overhead at 8 GPUs.
The reason for this is that TorchSWE maintains a large number of fields for each
simulated point, and issues different array operations on each field.
The amount of data needed for each element in the simulation does not allow 
the task granularity to be easily increased to the untraced Legion minimum of \~1ms per task, as each new
element added increases the memory footprint more than
it increases the average task granularity.
For such applications, leveraging tracing is a requirement, and \name{} enables
complex applications like TorchSWE to do so automatically.
TorchSWE itself contains enough task parallelism to hide communication latencies,
but needs tracing to first lower runtime overhead.
With \name{}, we are able to achieve between 0.91x--2.82x speedup on TorchSWE,
achieving nearly perfect scalability on 64 GPUs.

% \TODO{I have to transition into the 6500, 16 GPU anomaly and discuss it in a way that
% doesn't go into too much detail either.}

\subsection{Strong-Scaling}\label{sec:strong-scaling}

\begin{figure}
\centering
\includegraphics[width=0.9\linewidth]{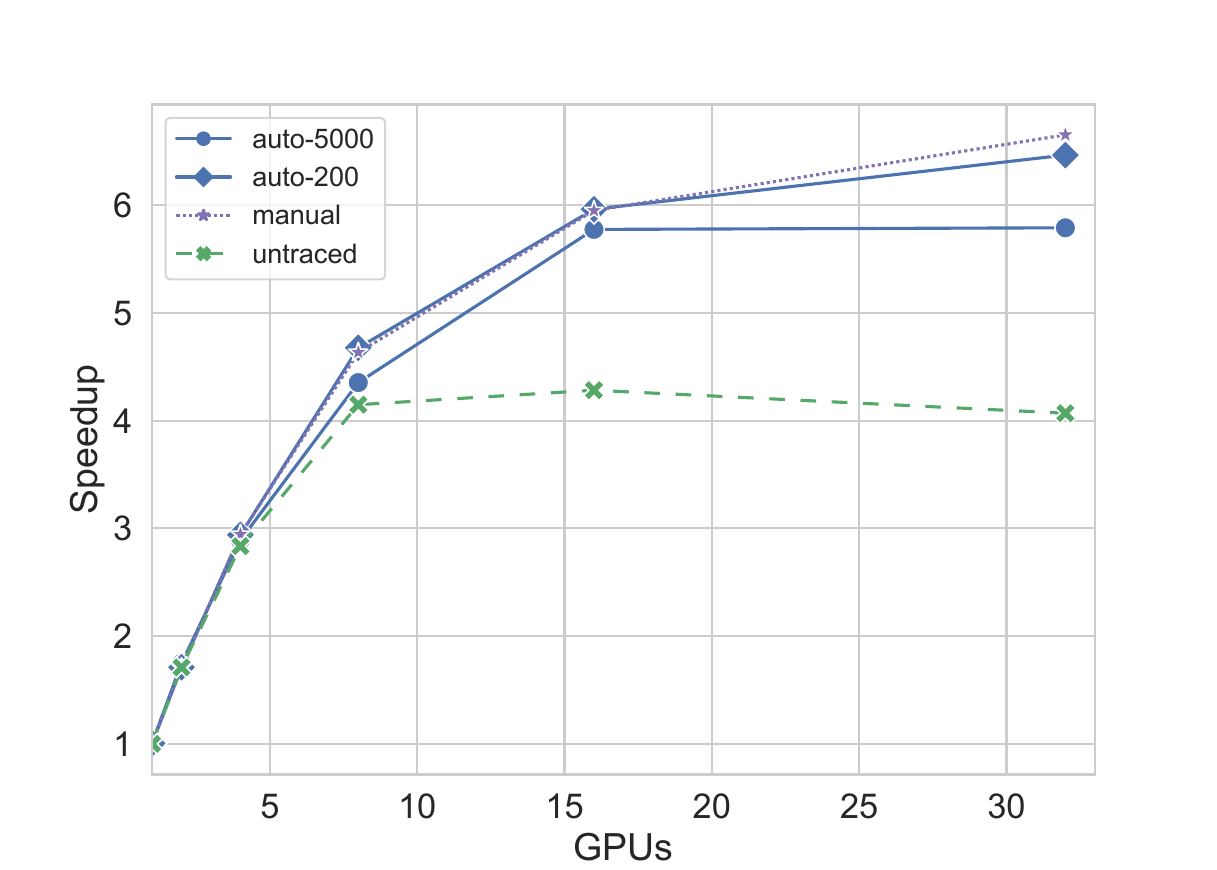}
\caption{Strong scaling of FlexFlow on Eos.}
\label{fig:flexflow-strong-scaling}
\end{figure}

We now move from scientific simulation codes to distributed deep neural
network training with FlexFlow~\cite{flexflow, unity}.
FlexFlow is a deep neural network framework that searches for hybrid
parallelization strategies for different layers of the network.
We perform a strong-scaling experiment with FlexFlow on Eos to train
the largest (\texttt{pilot1}) network from the CANDLE~\cite{candle} initiative\footnote{Due to engineering limitations in FlexFlow at the time of writing,
the network was parallelized only with data parallelism.}.
A strong-scaling study fixes the problem size on
a single processor, and increases the number of processors
while keeping total problem size constant.
To strong scale the training, we fix the batch size for single GPU, 
and then increases the number of GPUs available.

We compare the performance of FlexFlow with manual trace annotations, two
configurations of \name{} (discussed next), and no tracing.
As seen in \Cref{fig:flexflow-strong-scaling}, as FlexFlow scales up, 
the tasks become smaller and begin to expose Legion runtime overhead without
tracing, leading to slowdowns when scaling up.
The two configurations of \name{} differ in the maximum trace length
to be replayed (\name{}'s history buffer is the same, but recorded traces
are broken into pieces of a given maximum size).
The first (auto-5000) is the standard configuration with no maximum,
as used in all other experiments, and the second (auto-200) has a maximum
length of 200 tasks, which is similar to the length of the manually annotated trace.
As FlexFlow strong scales, the cost of Legion issuing the trace replay starts to
become exposed as the execution time of the trace decreases, leading to shorter
traces exposing less latency, and thus performing better\footnote{The Legion team is aware of this shortcoming and plans to address it in the future.}.
On 32 GPUs, the configuration of \name{} with a maximum trace length of 200 achieves between 0.97x
the performance of the manually traced FlexFlow, and achieves a 1.5x speedup
over the untraced FlexFlow.

\subsection{Overheads of \name{}}\label{sec:task-overhead}
We now discuss the overheads that
\name{} imposes over standard execution with Legion.
While we inherit the overheads of Legion's existing
tracing infrastructure~\cite{dynamic-tracing} (the cost
of memoizing traces), \name{} imposes two new sources
of overhead to measure: 1) the overhead on task launches
and 2) the time taken until a steady state is reached.

As discussed in \Cref{sec:trace-identification}, \name{}
intercepts the application's task launches and performs
some analysis work before forwarding the task launches
to Legion.
This analysis work includes launching asynchronous token buffer
processing jobs and manipulating traversals of the
trie data structures used for online trace identification.
To quantify this overhead, we ran a two node experiment on Perlmutter
and measured the time it took to launch (not analyze or execute)
Legion tasks with and without \name{} enabled.
We ran a two node experiment to ensure that the coordination
logic discussed in \Cref{sec:distribution} was included in timing.
We found that task launching took on average 7$\mu$s without \name{},
and on average 12$\mu$s with \name{}.
While \name{} increases the task launch overhead, this overhead
is still significantly lower than the amount of time
it takes to replay a task as part of a trace, which is ~100$\mu$s.
As such, the task launching cost of \name{} can still be effectively
hidden by the asynchronous runtime architecture.
The asynchronous analysis jobs that \name{} launches to process
task histories do not affect the critical path, and utilize Legion's
background worker threads.
While in theory these jobs could compete for the resources necessary for
Legion's dependence analysis, we have not yet encountered
an application where they caused a detriment in performance.

To measure the time taken until \name{} reaches a steady state
of replaying traces on our iterative applications, we report
the number of iterations until a steady state is reached.
\Cref{fig:steady-state-iterations} contains the iteration
counts needed for each application in \Cref{sec:weak-scaling}
and \Cref{sec:strong-scaling}, which range from 30 to 300.
These simulation and machine learning workloads would be
run in production for a significantly larger number of iterations,
so speedup in the steady state corresponds closely to end-to-end
speedup.
We note that the \cunumeric{} applications have a larger number
of required warmup iterations due to the dynamic behavior
discussed in \Cref{sec:motivating-example}, where a single
application-level iteration of the program does not 
necessarily correspond to a repeated sequence of tasks.

In terms of resource utilization, \name{} requires a modest amount
of CPU memory to store the history buffer of tasks for analysis.
\name{} runs the asynchronous string analysis (\Cref{sec:token-buffers})
on Legion's background worker threads.
We have not found these resource requirements to impact application performance or memory utilization.

\begin{figure}
\small
\begin{tabular}{|c|c|}
    \hline
    Application & Iterations Until Steady State \\
    \hline
    S3D & 50 \\
    HTR & 50 \\
    CFD & 300 \\
    TorchSWE & 300 \\
    FlexFlow & 30 \\
    \hline
\end{tabular}
\caption{Warmup iterations before \name{} reaches a replaying steady state.}
\label{fig:steady-state-iterations}
\end{figure}

\begin{figure}
\includegraphics[width=0.9\linewidth]{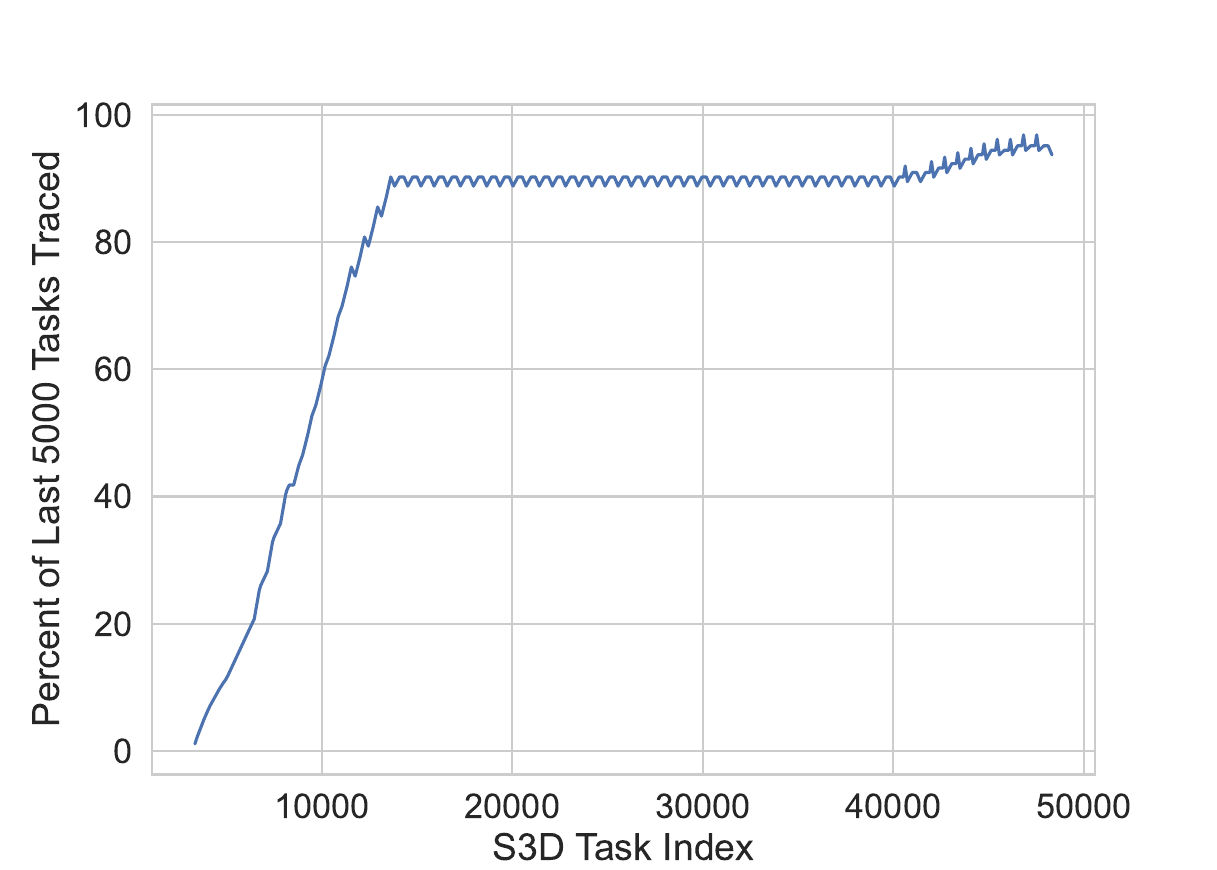}
\caption{Visualization of \name{} finding traces in S3D.}
\label{fig:search}
\end{figure}

\subsection{Trace Search}\label{sec:search-visualization}

To give intuition about the search process that \name{} performs,
we constructed a visualization of the amount of runtime overhead
that \name{} is removing over time.
\Cref{fig:search} is a visualization of S3D over time (for 70 iterations), where each for task launched by S3D, we display
how many of the previous 5000 tasks were traced.
For iterative computations, this procedure yields the expected
result, where \name{} spends time during program startup
discovering new traces, and then settles into a steady state.
The amount of traced operations increases slightly by the end of the execution, as \name{} finds a better set of traces that lowers the number of untraced operations.

\IGNORE{
\subsection{Separate Evaluation of String Search Algorithm?}

I'm not sure if we need this, but I can write some separate microbenchmarks to collect this.
}

\section{Related Work}\label{sec:related-work}

\paragraph{Just-In-Time Compilers}
Just-In-Time (JIT) compilers~\cite{self-jit, tracemonkeyjit, hotspotjit}
for dynamic languages have a tiered execution system,
where the target language is first translated to bytecode,
which is executed by an interpreter.
Frequently executed program fragments are then
compiled into native instructions for significantly
faster execution.
\name{} employs a similar architecture where a task-based
runtime system's dynamic analysis acts as the slow but
general interpreter,
and uses a tracing engine as the fast but specialized compiler.

Tracing-based JIT compilers such as TraceMonkey~\cite{trace-monkey}
record sequences of instructions executed at runtime
and generate optimized code for those sequences.
Method-based JIT compilers identify frequently invoked functions
in the target program and compile type-specialized
versions of those functions.
JIT compilers identify the desired instruction sequences or
methods to compile by relying on code landmarks like function
definitions and basic block addresses to maintain counters
of frequently executed program fragments.
Since \name{} views an unrolled stream of tasks, 
it must employ novel techniques for identification
of traceable program fragments.

JIT compilers also perform dynamic analysis to recover
data structures like call-graphs from the target program.
Sampling-based methods~\cite{calling-context-profiling} have been
developed to balance runtime cost of profiling each function call
with the accuracy of the sampled data structure.
Discovering traces in our work requires for long contiguous sequences
of issued tasks to be analyzed together, as a trace must
repeat several times to be considered by \name{}.
Breaking up these sequences with independent and non-contiguous samples can lead
to a loss of precision when discovering traces.
Instead, \name{} employs an always-on approach where all tasks are
analyzed, and uses a sub-sampling method on the
set of collected tasks to manage the trade off between
responsiveness of the trace analysis and length of the discovered traces.
An always-on approach is cheaper to use in the task-based
runtime system context than within a standard JIT compiler
as tasks are relatively coarse when compared to bytecode instructions.

\paragraph{Trace Caches}
Trace caches~\cite{trace-caches} have been used in processors
to improve instruction fetching bandwidth.
At a high level, trace caches record the common jump paths taken
through basic blocks, and pre-fetch those paths 
when revisiting the same basic blocks.
\name{} shares a similar architecture to trace caches, which
also use patterns in running programs to improve
the performance of a slower dynamic component (in this case, the
control-dependent instruction fetching).
Similarly to JIT compilers, trace caches also use landmarks
in executing programs to guide their decisions, which \name{}
is not able to exploit.
Also, by virtue of being implemented in hardware, the mechanisms
that trace caches must be simpler than the kinds of analyses
\name{} can use, which are implemented in software.

\paragraph{String Analysis}
\Cref{sec:token-buffers} contains a partial discussion of related string
analysis works---we continue the discussion here.
The most relevant string processing problem in
the bio-informatics community is \emph{motif finding}~\cite{motif-finding},
which is the problem of finding short (5--20 token long), fixed-length repeated
strings in a larger corpus.
The focus on a short and fixed sub-string length and a tendency to use genomic information
to guide the search makes these techniques not applicable to our problem.
Algorithms for document fingerprinting such as Moss~\cite{moss} have been developed
that accurately identify copies between documents.
In particular, these techniques are guaranteed to detect if
repetitions of at least a minimum size exist across documents.
Fingerprinting techniques are useful to detect whether there 
exist repeated sub-strings, but do not directly aid in finding the 
sub-strings themselves that have high coverage.

\paragraph{Inspector-Executor Frameworks}
\name{} is similar in spirit to Inspector-Executor (I/E) frameworks
that dynamically analyze program behavior and then perform optimizations~\cite{ie-1, ie-2}.
I/E frameworks generally focus on recording information
related to array accesses and use knowledge of these accesses
to perform compiler optimizations that parallelize or distributed loops.
In contrast, \name{} observes a dynamic sequence of tasks
and searches for repeated sub-sequences of tasks to record
as traces.

\paragraph{Task-Based Runtime Systems}

Several task-based runtime systems have been developed
for high performance computing~\cite{legion, starpu, parsec},
data science~\cite{dask, spark}, and 
machine learning~\cite{ray, pathways}.
One axis of runtime overhead that these different systems
impose on applications is the cost of dependence analysis.
The cost of dependence analysis is directly
related to the expressivity and flexibility of the 
runtime system's programming model.
Legion has an expressive data model that
supports \emph{content-based coherence}~\cite{legion-dep-analysis}, 
leading to a relatively expensive dependence analysis.
As a result, tracing~\cite{dynamic-tracing} was developed
to reduce the costs of the dependence analysis.
Both the StarPU~\cite{starpu-stf} and PARSEC~\cite{parsec-dtg}
runtime systems have modes that perform a dynamic dependence
analysis to extract parallelism, and these modes have been
shown to add overheads over the explicitly-parallel, analysis
free modes~\cite{task-bench}.
Tracing techniques could be applied within these runtime systems
to lower the overheads of the dynamic, implicitly parallel modes.

Techniques similar to tracing have also been developed
in other runtime systems to lower overheads.
A tracing-like technique called 
Execution Templates~\cite{omid-memoization} was developed
to cache control plane decisions in runtime systems for cloud-based
environments.
The Dask~\cite{dask} runtime system exposes an API for users to
explicitly construct and optimize task graphs~\cite{dask-compute-graph},
which is lower-level but more efficient than the standard individual
task launching API.
The Ray~\cite{ray} runtime system has recently added an execution mode
called ``Compiled Graphs''~\cite{ray-dag}, where users build explicit
computation graphs and issue them to Ray for lower overhead replay.
Finally, the CUDA runtime exposes a similar feature to tracing 
called CUDA Graphs~\cite{cuda-graphs}, where users may record a sequence of
CUDA kernel launches and replay the sequence with lower overheads.
Techniques used in \name{} could potentially be applied to these systems
to remove the requirement for users to be involved in the memoization
and optimization of these computations.

\section{Conclusion}

In this work, we introduce \name{}, a system and framework
for task-based runtime systems to automatically trace the
dependence analyses for repeated program.
By automatically detecting traces, \name{} is able to
improve programmer productivity by insulating programmers
against changing task granularity, and enable new applications
to take advantage of tracing.
We develop an implementation of \name{} that targets the Legion
runtime system and show that on the most complex Legion 
applications written to this date, \name{} is able to match the
performance of manually traced code, and effectively optimize
currently untraceable programs to improve the performance at scale by up to 2.82x.

\section*{Acknowledgements}

We thank Wonchan Lee, Manolis Papadakis and Shriram Jagannathan for their assistance
with Legate. We thank Seshu Yamajala for his assistance in running the S3D simulation.
We thank Elliot Slaughter for his assistance in debugging and running Regent programs.
We thank Mario Di Renzo and Caetano Melone for their assistance in running the HTR simulation.
We thank Zhihao Jia and Colin Unger for their assistance in running FlexFlow.
We thank Roshni Sahoo for her assistance in developing formal optimization problem in \Cref{sec:good-traces}.
We thank Danny Sleator and Sam Westrick for suggestions and pointers to related work around
the string analysis component of this work.
We thank Wei Wu for discussions about the PARSEC runtime system, and Cedric Augonnet
for discussions about the StarPU runtime system.
We thank (in no particular order) James Dong, AJ Root, Chris Gyurgyik, Rubens Lacouture, Shiv Sundram, Scott Kovach and Olivia Hsu for their discussions and feedback on this manuscript.
Rohan Yadav was supported by an NVIDIA Graduate Fellowship, and part of this work
was done while Rohan Yadav was an intern at NVIDIA Research.
This work was in part supported by the National Science Foundation under Grant CCF-2216964 and by Digital Futures at KTH Royal Institute of Technology. 
This research used resources of the National Energy Research
Scientific Computing Center, a DOE Office of Science User Facility
supported by the Office of Science of the U.S. Department of Energy
under Contract No. DE-AC02-05CH11231 using NERSC award
ASCR-ERCAP0026353.

\bibliographystyle{ACM-Reference-Format}
\balance
\bibliography{main}

\appendix
\section{Artifact Appendix}

%%%%%%%%%%%%%%%%%%%%%%%%%%%%%%%%%%%%%%%%%%%%%%%%%%%%%%%%%%%%%%%%%%%%%
\subsection{Abstract}

This artifact presents the computational artifact of Apophenia,
a system that automatically traces Legion applications.
This artifact is supported by an implementation of Apophenia
within the Legion runtime system and a standalone implementation
of the repeated sub-strings algorithm as described in the paper.
We also provide artifacts for the subset of our benchmarks that are
open-source.

We evaluated Apophenia on the Eos and Perlmutter supercomputers.
Each node of Eos is an NVIDIA DGX H100, containing 8 H100 GPUs with
80 GB of memory and a 112 core Intel Xeon Platinum.
Each node of Perlmutter contains 4 NVIDIA A100 GPUs with 40 GB of memory
and a 64 core AMD EPYC 7763.
Nodes of Eos are connected with an Infiniband interconnect, while Perlmutter
uses a Slingshot interconnect.
We compile Legion on Eos with the UCX networking module, and use the GASNet-EX
networking module on Perlmutter.

\subsection{Artifact check-list (meta-information)}

% {\em Obligatory. Use just a few informal keywords in all fields applicable to your artifacts
% and remove the rest. This information is needed to find appropriate reviewers and gradually 
% unify artifact meta information in Digital Libraries.}

{\small
\begin{itemize}
  \item {\bf Program: } A mixture of scientific and machine learning applications.
  \item {\bf Compilation: } C++ and CUDA compiler.
  \item {\bf Metrics: }
  Average throughput.
  \item {\bf Publicly available?: }
  Some aspects are publicly available, others are closed source.
  \item {\bf Archived (provide DOI)?: }
  Legion with Apophenia: \url{https://doi.org/10.5281/zenodo.13900083}.\\
  Repeated Substrings: \url{https://doi.org/10.5281/zenodo.13900514}.\\
  TorchSWE: \url{https://doi.org/10.5281/zenodo.13900751}.\\
  FlexFlow: \url{https://doi.org/10.5281/zenodo.13900858}.\\
\end{itemize}
}

%%%%%%%%%%%%%%%%%%%%%%%%%%%%%%%%%%%%%%%%%%%%%%%%%%%%%%%%%%%%%%%%%%%%%
\subsection{Description}

\subsubsection{How to access}

The version of Legion with Apophenia is available \href{https://gitlab.com/StanfordLegion/legion/-/tree/automatic-tracing/}{here}.
The standalone implementation of the repeated substrings algorithm is available \href{https://github.com/david-broman/matching-substrings}{here}.
The version of TorchSWE used for benchmarking is available \href{https://github.com/rohany/TorchSWE/tree/automatic-tracing}{here}, though executing it
with Apophenia requires a currently closed-source version of the Legate runtime and cuNumeric.
The version of FlexFlow used for benchmarking is available \href{https://github.com/rohany/FlexFlow/tree/automatic-tracing}{here}.

\subsubsection{Hardware dependencies}

Our experiments were run on server-class machines with multiple GPUs
per node. While these are not necessary,
the scaling and problem sizes that fit on each node will differ
on different setups.

\subsubsection{Software dependencies}

We run all our experiments with Python 3.11. Aside from that, a
standard supercomputer software stack (C++ compiler, CUDA installation, MPI installation) is expected.

%%%%%%%%%%%%%%%%%%%%%%%%%%%%%%%%%%%%%%%%%%%%%%%%%%%%%%%%%%%%%%%%%%%%%
\subsection{Installation}

Apophenia can be built and run with a standard Legion build,
using the version of Legion from the artifact. The exact
parameters to build Legion depend on the machine configuration.
A sample Legion build for an Infiniband-based cluster with
NVIDIA GPUs would invoke:
\begin{verbatim}
cd Legion/language
USE_CUDA=1 CONDUIT=ibv ./scripts/setup_env.sh
\end{verbatim}

Since FlexFlow is fully open-source, we also provide build instructions.
Using the provided version of Legion, FlexFlow can be built and
installed for an NVIDIA machine with
\begin{verbatim}
export FF_GPU_BACKEND="cuda"
conda create -n flexflow
source activate flexflow
conda install -c conda-forge cmake make pillow \
  cmake-build-extension pybind11 numpy pandas \
  keras-preprocessing onnx transformers>=4.31.0 \
  sentencepiece einops
conda install -c pytorch pytorch torchvision torchaudio
conda install rust
pip3 install tensorflow notebook
cd FlexFlow
mkdir build
cd build
../config/config.linux
make -j
\end{verbatim}

%%%%%%%%%%%%%%%%%%%%%%%%%%%%%%%%%%%%%%%%%%%%%%%%%%%%%%%%%%%%%%%%%%%%%
\subsection{Experiment workflow}

As a majority of our experiments are closed source, we do not
provide a script that can run the full experiment suite. We do
provide a command line that can be used to run the FlexFlow benchmark.
The given command line is intended for SLURM based clusters, but
additional configuration may be required depending on SLURM setup.

\begin{verbatim}
srun -N <NODES> \
  FlexFlow/build/examples/cpp/candle_uno/candle_uno \
  --warmup 30 \
  --batch-size 16384 \
  -ll:gpu <GPUS-PER-NODE> \
  -ll:fsize <GPU-MEM-IN-MBS> \
  -ll:util 4 \
  -ll:csize 30000 \
  -ll:zsize 5000 \
  -lg:enable_automatic_tracing \
  -lg:auto_trace:min_trace_length 25 \
  -lg:auto_trace:max_trace_length 200 \ 
  -lg:auto_trace:batchsize 5000 \
  -lg:auto_trace:identifier_algorithm \
      multi-scale \
  -lg:auto_trace:multi_scale_factor 500 \
  -lg:auto_trace:repeats_algorithm \
      quick_matching_of_substrings \
  -lg:inline_transitive_reduction \
  -lg:window 30000
\end{verbatim}

%%%%%%%%%%%%%%%%%%%%%%%%%%%%%%%%%%%%%%%%%%%%%%%%%%%%%%%%%%%%%%%%%%%%%
\subsection{Evaluation and expected results}

We expect that when used on our benchmark applications, Apophenia
finds and replays traces. On problem sizes that are Legion runtime-limited,
this should result in speedup.

%%%%%%%%%%%%%%%%%%%%%%%%%%%%%%%%%%%%%%%%%%%%%%%%%%%%%%%%%%%%%%%%%%%%%
\subsection{Experiment customization}

Apophenia exposes several runtime configurations that are accepted by
every Legion application for customizing the behavior.
\begin{enumerate}
    \item \texttt{-lg:enable\_automatic\_tracing}: enable automatic tracing.
    \item \texttt{-lg:auto\_trace:min\_trace\_length <N>}: minimum length trace to consider.
    \item \texttt{-lg:auto\_trace:max\_trace\_length <N>}: maximum length trace to replay.
    \item \texttt{-lg:auto\_trace:batchsize <N>}: size of the task history buffer.
    \item \texttt{-lg:auto\_trace:multi\_scale\_factor <N>}: minimum size of the adaptive analysis.
\end{enumerate}

\end{document}